# Protecting Spreadsheets against Fraud

*Roland T. Mittermeir* [1], *Markus Clermont* [2], *Karin Hodnigg* [1]

[1] *Institut für Informatik-Systeme*
*Universität Klagenfurt*
*AUSTRIA*
*Roland.Mittermeir@uni-klu.ac.at  Karin.Hodnigg@uni-klu.ac.at*

[2] *Software Quality Research Laboratory,*
*Department of Computer Science and Information Systems*
*University of Limerick*
*IRELAND*
*Markus.Clermont@ul.ie*

**ABSTRACT**

*Previous research on spreadsheet risks has predominantly focussed on errors inadvertently introduced by spreadsheet writers i.e. it focussed on the "end-user aspects" of spreadsheet development. When analyzing a faulty spreadsheet, one might not be able to determine whether a particular error (fault) has been made by mistake or with fraudulent intentions. However, the fences protecting against fraudulent errors have to be different from those shielding against inadvertent mistakes.*

*Faults resulting from errors committed inadvertently can be prevented* ab initio *by tools that notify the spreadsheet writer about potential problems whereas faults that are introduced on purpose have to be discovered by auditors without the cooperation of their originators. Even worse, some spreadsheet writers will do their best to conceal fraudulent parts of their spreadsheets from auditors. In this paper we survey the available means for fraud protection by contrasting approaches suitable for spreadsheets with those known from fraud protection for conventional software.*

## 1 MOTIVATION

The economic significance of spreadsheet errors is documented by the spectacular cases listed on the EuSpRIG web site [EuSpRIG]. The high frequency of errors of various kinds is documented by a host of empirical studies. An overview of these studies can be found in [Panko 00] or seen on Ray Panko's web site [Panko]. However, they could not reveal the motives underlying a particular error. Even in the study we did a while ago [Clermont et. al, 02] or in other un-documented assessments of sheets, we cannot state for sure that none of the deficiencies identified had a fraudulent background. We can only state that the way the original spreadsheet developers cooperated lets us assume that the majority of errors – if not all – were introduced without any hidden purpose.

Due to the Sarbanes-Oxley act, managers cannot rely on such well-meaning assumptions. Hence, research has to look deeper into the causes and consequences of spreadsheet errors. In this context it is important to note that risk prevention strategies always have to start at the head of the chain of problems to identify proper mechanisms to prevent them. For example, assuming that the kind of problems spreadsheet writers are dealing with remains of the same nature, investing in a person's spreadsheet specific education will hopefully have a positive impact on the error rates of the sheets this person writes[1]. This investment will, however, be without any effect, or possibly even have negative effects, if this particular employee is introducing errors on purpose.

To survey mechanisms protecting against the fraudulent introduction of faults or the fraudulent modifications of spreadsheets, one might review the general literature on software fraud considering the differences between spreadsheets and software developed by state of the art third generation languages and their respective professional development processes [Baskerville 93].

In the following sections we first consider general software crime and related protection mechanisms as well as the specific distinguishing features of spreadsheets. Based on these considerations, we will look at fraud protection mechanisms for spreadsheets. Unfortunately, we have to conclude that instantiating a comprehensive strategy still requires actions by both vendors of spreadsheet software and by the respective user organizations. Last but not least, it requires further research.

## 2 SOFTWARE CRIME

### 2.1 Strategies Dealing with Conventional Software

With conventional software systems, the distinction between program and data is clearly defined. In its weakest form, we have the compilation process as a universal frontier between software development, or its modification during maintenance, and the use of this software by having it operate on some data. Normally, users are distinct from developers and usually, software development is even further decoupled from its use, e.g. by distinguishing between development machines and production computers. Organizational precautions will ensure that there is only a one-way street from development to operation.

This distinction allows for a clear separation of the strategies protecting against frauds due to data manipulation and frauds due to software manipulation *per se*. Thus, frauds on the data level can be addressed by specific access rights. While fraudulent data manipulations can be made more difficult by the specific design of the data base schema or by incorporating cross checks in the application software, such strategies would have only a reduced effect were there to be no separation between developers and users.

The main strategy for assuring that software performs to its specification and only to its specification lies in the general quality assurance mechanisms of reviews and in testing. The advantage of reviews is that they can be made early in the development process. For fraud protection they will normally be executed in the form of inspections with item-lists specifically geared towards the identification of hidden functionality. The specifics of such lists are, as usual with inspections, language specific.

The systematic approach inherent in anti-fraud inspections has also to be used for anti-fraud testing. Among the vast spectrum of testing methods proposed, see [Zhu 97], one would specifically select functional testing methodologies. However, the problem is that one ought to test against the presence of code that is not required in the specification or – assuming the design has been adequately reviewed – not foreseen in the design of this particular software. To account for this, mutation testing [Murnane, Reed 01] might be an adequate approach.

---

[1] ) This positive assumption seems invalidated by [Panko, Sprague 97] claiming that the error rate is invariable with respect to seniority of the spreadsheet developers. However, this data is based on tasks of different complexity. Here, we assume risk specific education.

Finally, the interaction of data and software might also be used. Placing specific intermediate assertions will help to ensure that even intermediate results have to remain in correspondence with the conditions asserted. They are important in the fraud prevention context because persons operating with the system cannot use clever tricks – an alternative they might have even in the presence of openly visible checksums – to circumvent them. Assertions are embedded in code and therefore hidden from the user.

An aspect to be considered in conventional software that is less relevant for spreadsheets is that of hardware related frauds. The assumption is that spreadsheet users, be they also developers or only parametric users, have the sheet on their own PC. Hence, except for special cases, we can see almost no room for the use of encryption of data. Privacy aspects dealt with in the literature on security, specifically on the literature on data base security [Pernul et.al. 98] are relevant in general though. They are relevant concerning source data used in spreadsheet computations and in the transmission of spreadsheet results if those are to be used by persons other than the person operating a sheet.

**2.2 Spreadsheet Specifics**

Before checking to which extent the strategy "Lets do for spreadsheets what serves well for conventional software!" is used, one has to consider the differences between general software and spreadsheets.

A basic difference usually dwelt upon is that conventional software is produced by software professionals and spreadsheet software is developed by *end users*. Embarking too much on this difference would be dangerous though. It invites one equating end users with people of limited technical capability. This assumption could be quite misleading. Though spreadsheets can be written without too much in-depth knowledge of either programming or modelling, those that are subject to fraud are definitely not written by novices. One should rather assume that they are developed by personnel who, though being classified as end users, are well versed in both the domain they are working in and in spreadsheets themselves.

Spreadsheet writers differ from professional software developers, though, by following another development *process*. While 3GL software is usually created by following a rather structured development process based on some kind of specification, spreadsheets are usually written in a much more explorative fashion [Nardi, Miller, 1990].

This might be a consequence of the fact that, in contrast to spreadsheet development where developers unite all three roles, in the case of conventional software, the requirements owner and programmer are different individuals. Thus, even with relatively unstructured agile processes, specific steps have to be followed to allow for the coordinated effort of several people. Furthermore, even if the process is free from formal specifications, there are documents – whether in form of story cards [Beck 05] or in more conventional form is of secondary importance only – that can serve as reference points against which particular pieces of the system can be *verified*. With end user programming, i.e. whenever the requirements owner and the developer collapse into a single person, such reference documents seem to become unnecessary. Hence they exist only rarely. Even if the spreadsheet writer develops an explicit, written model of what is to be casted into the computations performed by the sheet and even if an independent design of the sheet has been made, these documents usually do not have the rigor of professional documentation.

Therefore, while the development of classical software will produce some *upstream documents* that can serve as reference points for the final code, this option is not available in spreadsheet development. Thus, the fact that the spreadsheet is written by a domain expert has severe direct and indirect consequences for fraud prevention strategies.

Besides these process-related aspects, one has to consider the *tool* itself. 3GL software is written with the support of some development tool, an editor that might be arbitratily independent of the language used. The formal text (source code) thus written is transformed in an automated step (compilation) to executable code. Only then is the compiled code applied to the data it is assumed to transform or to process. Spreadsheets, on the contrary, offer a 2(+1)-dimensional arrangement of cells such that each cell can be either empty, contain a constant value, or a formula. Constant values can be considered to be either part of the program, serving as constants or labels, or as data to be processed, which would be called 'input' if we were discussing conventional, 3GL programs.

The analogy of output is missing insofar as each cell that has some contents displays its value, irrespective of whether this value is to be assumed as result of the spreadsheet model or whether it is just an intermediate. Methodologies prescribe modularizing sheets such that distinct input-, computation-, and output areas appear on the sheet [Ronen et al., 89], [Burnett, 01]. However, these methodologies are observed only to a limited extent and one may also raise arguments against their adoption.

This *coupling of program with data* has, however, severe consequences for quality assurance and authentication procedures. Certainly, one could test spreadsheets in the same way as conventional programs are tested. In the conventional case, however, the tester is free from the risk of inadvertently changing the program; with spreadsheets, this risk can only be controlled by cell protection and/or locking mechanisms [Burnett et. al., 01].

## 3 PROTECTION MECHANISMS FOR SPREADSHEETS

In this section we discuss mechanisms for protecting spreadsheets against errors introduced in a fraudulent manner. Before delving deeper into this issue, it is important to say that fraud is a socio-psychological phenomenon with economic consequences. Hence, as with other social or psychological phenomena, technical means alone are insufficient and technical or methodological means cannot prevent fraud. They can only establish barriers that are difficult to overcome. Ingenious people, however, will always be smart enough either to overcome or to circumvent such barriers. Hence, *organizational means* are necessary to embed the recommendations discussed below in an appropriate, organization-specific manner to avoid having malicious users being able to circumvent what is placed in front of them to protect the integrity of models and computations.

Having said so, we discuss in the following sections how reviewing, testing, the placement of assertions, and authentication – techniques taken from general software fraud protection – can be used to protect spreadsheets against the incorporation of fraud-permissive code or against fraudulent modifications.

### 3.1 Reviews and Inspections

In conventional software development, inspections, though still not universally used in practice, are attributed to be the most effective error detection process steps [Endres, Rombach 03, Rombach et.al., 02]. Interestingly, not all software in use has been subjected to a review while nobody would trust an untested piece of software. The fact that people trust more in the less strong testing technology than in reviews should be an important point to consider when devising a comprehensive fault prevention strategy. It also constitutes an interesting research question in its own right.

There are several forms of reviews, distinguished by their purpose, position in the development cycle and the particular techniques applied. The techniques range from walkthroughs to inspections. While walkthroughs amount essentially to a manual execution of the programme in a particular group setting, inspections usually check code linearly against specific deficiencies. The deficiencies are taken from a list that contains items that are known to be particularly error prone within the given development context. However, one might follow different reading strategies [Biffl, 01]. Reviewing development products from pre-coding phases normally requires a specific reading strategy that takes care of the non-linearity of the contents of the respective document. In any case, one has the requirement that the full document is checked and the reading strategy has to ensure that nothing is missed.

When checking spreadsheets, one has first to note that they do not share the linearity of 3GL-programs. The arrangement of cells is two-dimensional on the sheet and if there are references between different sheets of a work-book, one might consider a spreadsheet as three-dimensional arrangement of interrelated cells. Hence, particular recommendations for a reading strategy are needed. Of course, one might follow a line-by-line, column-by-column strategy. This will ensure that each cell is checked, a requirement called for by most recommendations [Rothermel, 01; Sajaniemi, 98; Burnett, 01]. However, one might doubt whether the high number of cells present in professional spreadsheets makes this strategy productive. Cell references can stretch over arbitrary distances and hence this pseudo-linear reading might miss important semantic relationships, relationships that were placed (or missed) with fraudulent intentions.

Further, the fact that many spreadsheets contain substantial portions of a repetitive nature must be considered. If ignored, reviewers may become bored with a resulting loss in care and attention. However, becoming bored or tired is a problem for *any* kind of reviews. Therefore, irrespective of the special review strategies, reviewing methodologies call for strict time limits to be observed in review sessions.

The work of the Klagenfurt group proposes strategies to lessen this effect. In [Ayalew et al., 00; Clermont 03], logical areas are proposed. They group cells of identical content into sets that can then be checked as sets on their defining basis. Originally, this concept was developed to identify differences between regularity in the layout and regularity in the semantic content of a cell. An empirical evaluation [Clermont et. al, 02] has shown that the related tool for supporting a reviewer not only has a high detective power – even in sheets that were considered correct, the reviewer found error rates that are in line with those reported in the literature [Panko, 98] – it also dramatically lessens the time needed for performing the inspections.

This effect is to be expected by comparing the approach to other software certification strategies. In conventional software development, (full) path testing is accepted as the most thorough coverage strategy. In practice, full path testing is infeasible though [Weyuker 86]. Hence, one embarks on less rigorous strategies such as simple path tests where for each loop at least one case that executes the body and one case that misses the loop body are run.

With spreadsheets, we assume that the analogues of loops are cells with content copied from other cells. The finiteness of the (2+1) dimensions of the sheet would allow an exhaustive assessment of each cell. However, performing just a check of each logical area amounts exactly to what is checked with a simple path test. Moreover, since the areas are identified automatically, even minor differences in arbitrarily complex cell contents will be identified. Thus, if fraudulent modifications lead to irregularity in structures, reviewers are specifically pointed to such spots. If fraudulent modifications lead to consistent modification, the reduced number of items to be considered improves the chance that they do not slip the attention of the reviewers.

The concept of grouping repetitive structures for review purposes can be extended to groups of cells that are defined according to a repetitive pattern. With the so defined semantic classes [Mittermeir, Clermont 02], even higher levels of aggregation can be safely checked for fraudulent deviations.

These concepts have been formally defined in [Clermont, 03] where an orthogonal concept based on the data flow between semantic units is additionally defined. With this, a reviewing discipline is proposed that is (or strives to be) in accordance with the application model formulated by the spreadsheet. How the concepts just described can be combined for auditing has been described in [Clermont, Mittermeir, 03]. There, three specific reading strategies for spreadsheets are proposed.

It should be mentioned here that in searching for fraud, several rounds of passing through a sheet are necessary and these rounds might follow different sheet traversal strategies [Biffl, Halling, 03]. While a check for accessing files that do not belong to the legitimate interface might be done automatically by a line-wise or column-wise check of each cell (including hidden cells!), auditing the algorithm itself might instead require a trace along or against the flow of data. A rather selective strategy might be to check for cases where a choice of misleading labels can lead to effects that are to be prevented.

### 3.2. Testing

Research on testing conventional software has produced a host of literature (see [Zhu 97, Hierons 02] for an overview). Rothermel and his co-authors proposed a specific testing methodology for spreadsheets [Rothermel 01].

Given Dijkstra's statement that testing can only show the presence of faults, but never their absence [Dijkstra 72], the high emphasis on testing might be surprising. Dijkstra's remark applies in an analogous manner to fraudulent deviations of a program from what should be the program following its specification. In fact, testing is weaker than either proofs or checking against specifications in a review. The reason for this is that the latter two deal with a program in its intentional form while testing deals only with a sample taken from the possible extensions of input-output transformations to be performed by a program. Hence, while testing can only raise the level of confidence one can have in the correctness of a program, its fault detection characteristics are sufficiently different from those of reviews or proofs so that any solid quality management strategy would employ a mix of these quality assurance strategies [Mittermeir, 04].

A basic assumption of testing is that it is performed after coding is completed. In industrial software production, the separation of the coding and testing phases is even emphasised by having different personnel for programming and for testing and clear prescriptions on the quality a module has to have before it is to be passed to the testing group and subjected to formal testing. Configuration management systems and special conventions are needed to ensure that there is no conflict of versions between testing and correcting.

Due to the integration of data and code, such assumptions seem problematic with spreadsheets. Any change of (input-) data might, by mistake, lead to a change in the program itself. Further, manual instrumentation of (normal) spreadsheets seems impossible, since such changes would distort not only the internal structure of the sheet, but also its external appearance. Cell protection might be an answer to the risk of inadvertently changing a cell that belongs to the "program part" rather than to the "input part" of the sheet. Doing so on a discretionary basis might itself be error prone if one aims at testing to raise the confidence in the functional correctness of the sheet. If one plans to use testing as fraud identification strategy, however, it would appear to be totally insufficient.

To improve confidence in testing as a fraud identification strategy, two changes seem mandatory:
a) the physical separation of data from code,
b) the temporal separation of data changes from changes in code.

The first requirement is approached by having methods ensuring firstly specific regions of the sheet for each one of the three categories: input cells, "code", and cells bearing final results. Secondly, block-wise protection locks are needed for the latter two categories of cells. To provide separation with a higher level of protection from the system would, however, require a methodology where workbooks are structured in data-sheets and program sheets such that the front sheet shows only input values and the final result values. Computations would then take place on a different sheet (or possibly on a stack of different sheets) of the workbook.

This separation will require some discipline from the developers, but it does not really raise the overall complexity. In contrast to other proposals that recommend separation of formulas from input cells, here the lower level sheets can be prepared in exactly the manner in which they would have been prepared anyway. Even the initial desk checks might be done by the original spreadsheet writer as she or he used to do. Only before submitting the sheet to testing, all values in intended "input cells" have to be replaced by a reference to the respective cell on the front sheet and, for all "output cells", the front sheet has to get a simple reference to the respective "program sheet". Labelling information still has to be copied to the "front sheet" containing the value view of this spreadsheet model. This could, however, be done with tool support if one follows labelling identification strategies such as proposed by [Hipfl, 04].

However, in the fraud prevention case, the crucial issue remains whether the sheet that has been tested will also be the sheet that is actually used in operation. To achieve this, one would need to totally and finally remove the code modification capability from the person operating the sheet. The OpenOffice extension TellTable [Nash et al, 04] supports this strategy by placing the real spreadsheet on a protected server. Users operate on a view of the sheet presented via a VNC client. Thus, they are allowed to perform parametric variation of the sheet while the algorithmic core is protected and can be changed only by a special group of users. This separation of interface from internals of the sheet would also enhance auditability of change traces. With tools supporting focussed analysis of modification histories [Baxter, 04], auditors can efficiently ensure that the sheets in operation are still conformant to those that have been certified previously.

Another answer to this issue would be to use compiled spreadsheets. Compiled spreadsheets seem to be not really an additional burden on the spreadsheet development process. The spreadsheet compiler would just clearly define the boundary between development and operation. In fact, the compiled spreadsheet would, from the user's point of view, just consist of the "front sheet" proposed above. However, this front sheet is not manually constructed but prepared by the spreadsheet compiler. Underneath this front sheet, there can be arbitrarily optimized code that might be protected in a highly professional manner.

It is to be seen that the advantage of using a spreadsheet compiler over the use of protected cells might be limited for the initial version of a spreadsheet. However, those sheets that are fraud-critical usually have a long lifespan during which they get modified. With the compiler solution, a trace history over the various versions of such sheets has the compilation run as a clearly defined event that marks the creation of a new version.

A word of caution still has to be said. As with searching for errors, searching for frauds is exploring the open universe against the finite space of the specified (or intended) solution. However, in seeking errors, one has a rich empirical basis for the kinds of errors humans commit. In the case of seeking for fraudulent deviations from the specified (or intended) solution, this experience base is very limited. Hence, the identification of appropriate test data sets, which basically amounts to a very carefully designed sampling strategy, is much more difficult and laden with risk.

### 3.3. Assertions

Assertions are a strategy for relating program code to its specification and, if the assertions are to be executed during operation, for checking whether the state defined by the operation of the program on a particular stream of input data conforms to the specification.

With spreadsheets, assertions play a limited role if one is not prepared to accept checksums as a weak proxy of such assertions. However, we see in the work of Ayalew [Ayalew 01, Ayalew et al, 00] a certain approximation to assertions that might have some merit in fraud prevention. His idea was to have a layer underneath the normal spreadsheet on which the ranges to be assumed by the values of selected cells can be specified. Using interval arithmetic, the specified ranges can then be compared with the ranges resulting from executing the functions and operations of the original spreadsheets on the ranges of cells higher up in the data flow. This allows establishing inclusion relationships between the actual value, the computed range, and the expected range of result cells. Deviations will point to potential faults in either the spreadsheet writer's model or in its implementation in the original spreadsheet.

While the direct transfer of this idea to fraud prevention might be of limited use, its generalization seems highly promising. It offers a layer underneath the spreadsheet that is under the application expert's control and that checks certain spots of the sheet in a way that is not directly accessible to the application expert operating the sheet. Thus, modifications of the original sheet will be possible as long as these modifications remain within the competencies granted to this person. Obviously, this also requires additional protection mechanisms as discussed in connection with a spreadsheet compiler.

### 3.4. Authentification

Finally, one has to point to classical security mechanisms such as authentification or water marking ([Craver et al., 98], [Collberg et. al., 99]). Both rest on cryptographic tech-niques and would, to be of practical use, require compiled sheets or at least a strong physical separation between data and code.

The idea is basically that the content of the program portion of the sheet is subjected to some cryptographic encoding such that any modification in the program would lead to a different cipher. Consequently, the program is sealed against unauthorised changes and users of this program can be sure that they used the program they intended to use and not some slightly modified clone of it.

Since adequate treatment of this topic would extend the scope of this paper, we refer to the related literature for cryptographic protection of conventional software (c.f. [Devanbu, Stubblebine, 00]). Here, the differences between conventional software and spreadsheets will not matter as long as one can clearly separate code from data. It is to be understood, however, that applying any of the techniques discussed above, regardless of how efficient or effective they might be for dealing with inadvertent errors, will be wasted effort as long as one cannot be sure that the sheet that was operated was the sheet that has been previously examined.

## 4. SUMMARY

Though spreadsheets have to be understood as programs written in a very high level programming language, mechanisms to protect them against fraud have to recognize the substantial differences between them and classical software. Key to these differences is that the multi-person process, which is inherent in conventional software production and has a clear differentiation of roles, is lost in favour of a single person process. The latter is apparently much harder to control from the outside. Intermediate documents that might otherwise serve as reference points are usually missing. To compensate for this, organizational measures have to be put in place. Likewise, one has to take care of the fact that spreadsheet systems work in an interpretative mode and that data and program is not necessarily separated on spreadsheets.

Separation of data and "program" will eventually be a key requirement when one wants to protect spreadsheets against fraudulent modifications, but we must be aware that this would radically change the way spreadsheets are used and developed. Up to now, the approaches suggesting separating input data from programs proper have, in practice, not been blessed with high acceptance.

Inspections of various sorts might help to check sheets. Techniques that have analogous effects to loop aggregation might improve the efficiency and even the effectiveness of such reviews and testing approaches will also have their place in fraud prevention. However, all effort invested will be futile as long as one cannot be sure that the spreadsheet assessed will be the spreadsheet executed. This assurance can be given by cryptographic methods. However, they require a clear identification of what is code and what is data.

**Acknowledgement**
The authors would like to thank Dr. Tom Arbuckle and the anonymous referees for helpful comments on a draft of this paper.